\documentclass[aa]{aa} % for the letters 
\usepackage{graphicx}

\usepackage{txfonts}
\usepackage{hyperref}
\usepackage[dvipsnames]{xcolor}

\newcommand{\sersic}{S\'ersic\xspace}

\newcommand{\prospector}{\textsc{Prospector}\xspace}
\newcommand{\forcepho}{\textsc{ForcePho}\xspace}
\newcommand{\sig}{\ensuremath{\sigma}\xspace}

\newcommand{\Mstar}{\ensuremath{M_\star}\xspace}

\let\oldAA\AA
\renewcommand{\AA}{\text{\oldAA}\xspace}

\newcommand{\otwolong}{\ensuremath{\mathrm{[OII]}\lambda\lambda 3726,29}\xspace}

\newcommand{\otwo}{\ensuremath{\mathrm{[OII]}}\xspace}

\newcommand{\cthree}{\ensuremath{\mathrm{CIII]}\lambda\lambda 1907,09}\xspace}

\newcommand{\othreelong}{\ensuremath{\mathrm{[OIII]}\lambda \lambda 4959,5007}\xspace}

\newcommand{\othree}{\ensuremath{\mathrm{[OIII]}}\xspace}

\newcommand{\gal}{\text{JADES-GS8-RL-1}\xspace}

\newcommand{\be}{\ensuremath{\beta_{\rm UV}}\xspace}

\newcommand{\fesc}{\ensuremath{\mathrm{f_{esc}}}\xspace}
\begin{document}

   \title{Zapped then napped? A rapidly quenched remnant leaker candidate with a steep spectroscopic \be slope at z=8.5}
   \titlerunning{Zapped then Napped}

   %\subtitle{I. Overviewing the $\kappa$-mechanism}
   \authorrunning{Baker et al.}

\author{William M. Baker
          \inst{1, 2, 3}\fnmsep\thanks{william.baker@nbi.ku.dk}
          \and
          Francesco D'Eugenio \inst{2,3}
          \and 
          Roberto Maiolino \inst{2,3,4}
            \and
          Andrew J. Bunker \inst{5}
            \and
            Charlotte Simmonds \inst{2,3}
            \and
            Sandro Tacchella \inst{2,3}
            \and
            Joris Witstok \inst{6,7}
            \and
            Santiago Arribas \inst{8}
            \and
            Stefano Carniani \inst{9}
            \and
            St\'ephane Charlot \inst{10}
            \and
            Jacopo Chevallard \inst{4}
            \and 
            Mirko Curti \inst{11}
            \and
            Emma Curtis-Lake \inst{12}
            \and
            Gareth C. Jones \inst{2,3}
            \and
            Nimisha Kumari \inst{13}
            \and
            Pierluigi Rinaldi \inst{14}
            \and
            Brant Robertson \inst{15}
            \and
            Christina C. Williams \inst{16}
            \and
            Chris Willott \inst{17}
            \and
            Yongda Zhu \inst{14}
          }
          \institute{DARK, Niels Bohr Institute, University of Copenhagen, Jagtvej 155A, DK-2200 Copenhagen, Denmark
              %\email{william.baker@nbi.ku.dk}
        \and
        Kavli Institute for Cosmology, University of Cambridge, Madingley Road, Cambridge, CB3 OHA, UK
        \and
        Cavendish Laboratory - Astrophysics Group, University of Cambridge, 19 JJ Thomson Avenue, Cambridge, CB3 OHE, UK
        \and
        Department of Physics and Astronomy, University College London, Gower Street, London WC1E 6BT, UK
        \and
        Department of Physics, University of Oxford, Denys Wilkinson Building, Keble Road, Oxford, OX1 3RH, UK
        \and
        Cosmic Dawn Center (DAWN), Copenhagen, Denmark
        \and
        Niels Bohr Institute, University of Copenhagen, Jagtvej 128, DK-2200, Copenhagen, Denmark
        \and 
        Centro de Astrobiolog\'ia (CAB), CSIC–INTA, Cra. de Ajalvir Km.~4, 28850- Torrej\'on de Ardoz, Madrid, Spain
        \and
        Scuola Normale Superiore, Piazza dei Cavalieri 7, I-56126 Pisa, Italy
        \and
        Sorbonne Universit\'e, CNRS, UMR 7095, Institut d'Astrophysique de Paris, 98 bis bd Arago, 75014 Paris, France
        \and
        European Southern Observatory, Karl-Schwarzschild-Strasse 2, 85748 Garching, Germany
        \and
        Centre for Astrophysics Research, Department of Physics, Astronomy and Mathematics, University of Hertfordshire, Hatfield AL10 9AB, UK
        \and
        AURA for European Space Agency, Space Telescope Science Institute, 3700 San Martin Drive. Baltimore, MD, 21210
        \and
        Steward Observatory, University of Arizona, 933 North Cherry Avenue, Tucson, AZ 85721, USA
        \and
        Department of Astronomy and Astrophysics, University of California, Santa Cruz, 1156 High Street, Santa Cruz, CA 95064, USA
        \and
        NSF National Optical-Infrared Astronomy Research Laboratory, 950 North Cherry Avenue, Tucson, AZ 85719, USA
        \and
        NRC Herzberg, 5071 West Saanich Rd, Victoria, BC V9E 2E7, Canada
        }

   %\date{Received ; accepted }

% \abstract{}{}{}{}{} 
% 5 {} token are mandatory
 
  \abstract
  {We use NIRSpec MSA spectroscopy and NIRCam photometry to explore the properties of \gal, a rapidly quenched, $z=8.5$ galaxy with a stellar mass of $\rm 10^{8.9}M_\odot$, a steep blue UV slope, a Balmer break, and no sign of strong emission lines.
  With a \be=-2.8$\pm 0.2$, as measured from the NIRSpec spectrum, \gal is consistent with negligible dust attenuation and little to no contribution from the nebular continuum alongside a probable high escape fraction. The \be slope measured from photometry varies from -3.0 in the central regions to -2.2 at the outskirts suggesting possible regional differences in the escape fraction.
  There are no high-ionisation emission lines, only a tentative 2.9\sig detection of \otwolong. Using photometry, this emission appears to be extended, possibly corresponding to weakly ionised gas expelled during or after the quenching process. \gal is spatially resolved with a half-light radius of 240 pc and has an exponential, disc-like morphology.
  It appears to have formed all its stars in a short burst within the past 100 Myr with a formation time of $\approx$70 Myr and a quenching time of $\approx$30 Myr. This quenching would have occurred rapidly, making it a more distant example of the kind of low-mass ``mini-quenched'' galaxies previously observed at high-z. Due to the extremely blue \be slope, our best-fit model predicts a high value for \fesc of >10\%, consistent with the value derived from the \be slope, which when combined with our extraordinarily low O32 upper limit suggests \gal is a fascinating example of a high-z ``remnant leaker'' in one of its earliest phases, deep in the epoch of reionisation.}
  % % context heading (optional)
  % % {} leave it empty if necessary  


   \maketitle
%
%-------------------------------------------------------------------

\section{Introduction}
\label{s.intro}

In the early Universe, gas accretion onto galaxies is believed to be more stochastic -- resulting in "burstiness" in galaxy star-formation rates \citep{Dekel2006,Hopkins2014, Sparre2017, Sun2023, Mason2023}. This is believed to particularly affect low-mass galaxies \citep{Tacchella2023b,Endsley2024, Endsley2024b, Langeroodi2024}, which are expected to go through burst phases, between which they remain dormant until the latest accretion of new gas to reignite star formation \citep{Strait2023,Looser2024,Looser2023, Dome2024}. This is therefore thought to be a temporary phenomenon, unlike high-redshift massive galaxy quenching \citep[e.g.][]{Carnall2024,deGraaff2024, Weibel2024,Baker2024c} which are dominated by strong Balmer breaks \citep[and sometimes even 4000-\AA breaks][]{Glazebrook2024, Baker2024c} and extended periods of quiescence.

JWST has uncovered examples of the former so-called "mini-quenched" galaxies both spectroscopically \citep{Strait2023,Looser2024} and photometrically \citep{Trussler2024}. 
These galaxies are characterised by Balmer breaks, a sign of an evolved (10-100\;Myr old) stellar population, combined with a relatively steep UV slope (\be$\approx-2$), a sign of recent star formation (50--100 Myr), but do not display prominent emission lines \citep[see e.g.][]{Looser2024} which trace star formation on time-scales shorter than 10~Myr \citep[e.g.][]{Kennicutt2012}. 

A steep \be ties into many properties of a galaxy. It requires an extreme lack of both dust and nebular continuum as the presence of either can significantly flatten \be.
The intrinsic slope is itself set by the stellar population properties of the galaxy, such as stellar-population age, metallicity, and the amount of nebular continuum emission \citep{Wilkins2011,Topping2024}.
High-z steep \be slopes have been reported previously as part of samples \citep[e.g.][]{Furtak2023, Cullen2023,Topping2024, Heintz2024, Saxena2024, Dottorini2024}, but on average the \be slopes at high-z are not observed to have extreme values \citep{Roberts-Borsani2024}.
In addition, these few observed steep \be slope galaxies are normally found to have strong emission lines, as found, for example, in \citet{Saxena2024} and for the stack in \citet{Dottorini2024}.

High-z galaxies with steep \be slopes are also important because they are thought to be key players in the Epoch of Reionisation \citep[EoR, e.g.][]{Robertson2022, Simmonds2024a} due to the steep \be slopes requiring high escape fractions (\fesc) \citep[as found directly in the low-redshift Universe, e.g.][]{Chisholm2022}. The EoR is the period within which the neutral intergalactic-medium (IGM) became ionised and which likely ended around redshifts 5.3-7 \citep{Fan2006,Bosman2022, Zhu2024}.

The idea is that galaxies can be split into two types of Lyman Continuum leakers depending on their properties \citep{Katz2023, Simmonds2024}. 
The first are "bursty leakers" with increasing star-formation histories ($\rm SFR_{10Myr}/SFR_{100Myr}$>1), alongside high-ionisation parameters, whilst the second are the remnant leakers, which are characterised by decreasing star-formation histories ($\rm SFR_{10Myr}/SFR_{100Myr}$<1) and lower ionisation parameters. Remnant leakers have had a previous extreme burst of star-formation that is greater than their current star-formation, which has likely been shutdown most recently due to the effects of feedback. The idea is that this feedback, as well as halting star-formation, may also have enabled a greater escape fraction of Lyman continuum photons due to the creation of channels out of the ISM, through which these photons can travel \citep{Zackrisson2013}. Remnant leakers are also believed to show properties similar to density-bounded nebulae \citep[i.e. ionising photons escape due to there being no more gas available to ionise, e.g.][]{McClymont2024}, compared to the bursty leakers which show properties of ionisation-bounded nebulae.

This ties into UV slopes because in order to get a steep UV slope, one requires not only to be almost dust-free, but also to be both metal-poor and with little contribution from the nebular continuum. 
This lack of nebular continuum is also a strong sign of a remnant leaker.

However, a key factor here is exploring timescales. Depending on which phase of a SFH is observed greatly affects what kind of spectrum you will see. This is important for UV slopes, Balmer breaks, and burstiness. Galaxies can be observed in burst phases \citep{Looser2023, Endsley2024, Langeroodi2024}, mini-quenched phases \citep{Looser2024}, or anything in between or after \citep[such as rejuvenation, e.g.][]{Witten2024}.

Galaxies with unexpectedly steep \be slopes and bright luminosities appear to be frequently observed at very high redshift and are often referred to as "Blue Monsters" \citep{Ferrara2024} 
The idea behind blue monsters is based on an attenuation-free model \citep[AFM,][]{Ferrara2024b}, where at high-z there would be an almost complete lack of dust attenuation, which could be due to processes such as radiation driven outflows of dust and gas within very early galaxies \citep{Ferrara2024b}. This outflow would make the galaxy very blue (due to lack of dust) but also rapidly quenched due to the gas outflow. 
These blue monsters may explain the unexpectedly luminous galaxies observed at $z\gtrsim10$, which otherwise challenge both theoretical models and observations at lower $z$ \citep[e.g. ][]{Naidu2022b,Curtis-Lake2023, Finkelstein2022,Castellano2024,Carniani2024, Ferrara2023}
%, via an attenuation-free model \citep[AFM,][]{Ferrara2024b} whereby at those redshifts there is almost a complete lack of dust attenuation, which could be due to processes such as radiation driven outflows of dust within these very early galaxies \citep{Ferrara2024b}. These "blue monsters" would then undergo rapid/mini quenching into a "lull" phase before reigniting star formation. }

In this paper, we will explore \gal, whose unique combination of continuum and emission-line properties suggest an in-between evolutionary stage, with a steep \be slope as predicted for blue monsters, but also with a Balmer break typical of mini-quenched galaxies. 

Throughout this paper, we use the \citet{PlanckCollaboration2020} cosmological parameters alongside a \citet{Chabrier2003} Initial Mass Function (IMF).

\begin{figure}
   \centering
    \includegraphics[width=1.0\columnwidth]{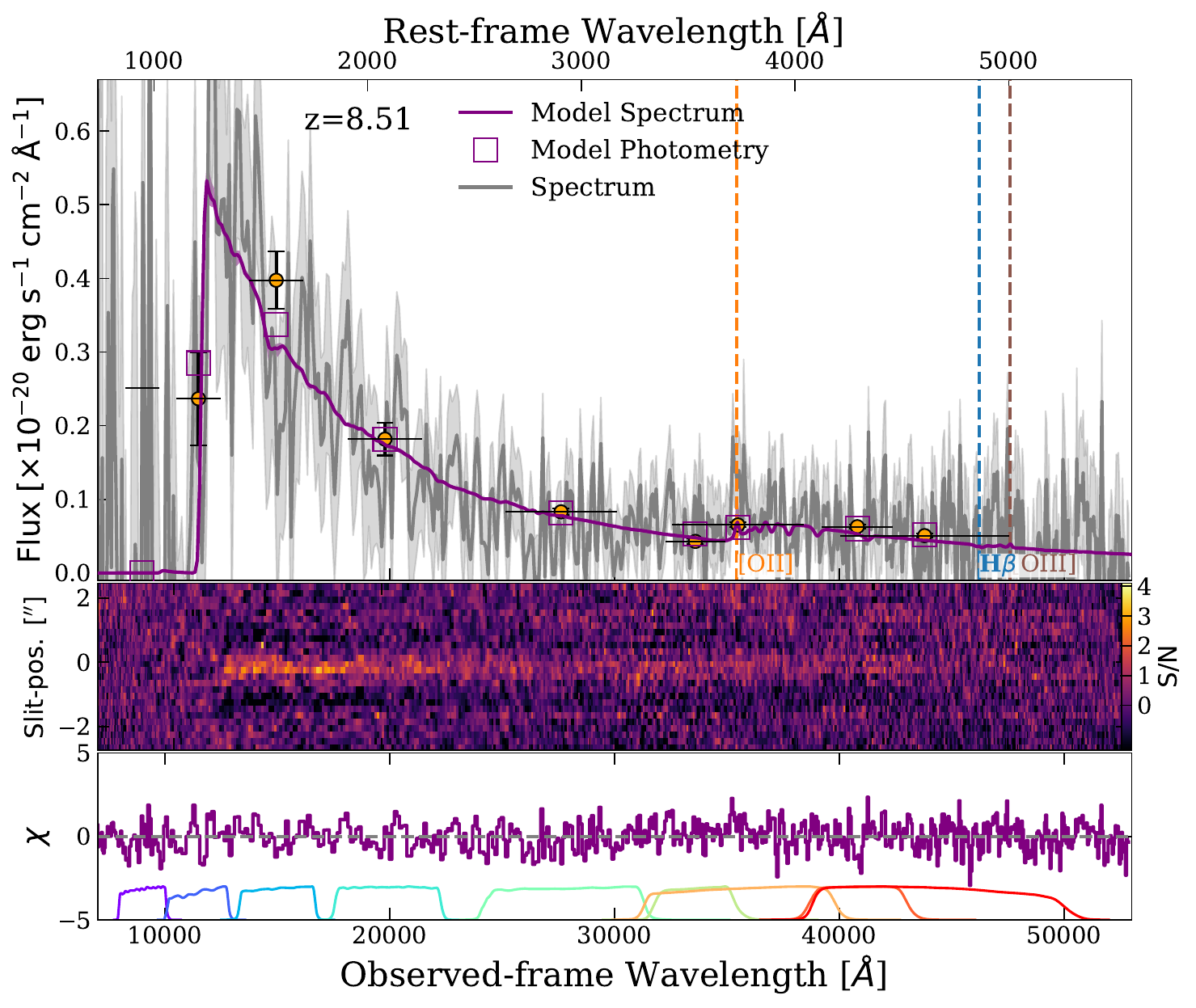}
   \caption{Upper panel, 1D spectrum (grey), photometry (yellow points), and best fit spectrum and photometry (purple) for \gal. We see evidence for a steep \be slope, with no strong emission lines. The spectrum shows a clear Lyman $\alpha$ break, a Balmer break, and tentative \otwo emission at the expected wavelengths. Common emission lines are overplotted as dashed lines. 
   Middle panel: 2D spectrum of slit position versus wavelength colour coded by signal to noise ratio. 
   The normalised fit residuals (bottom panel) show no high-order modulation with wavelength. The NIRCam filter curves are overplotted.
   }\label{fig:Spectrum}
\end{figure}

%--------------------------------------------------------------------
\section{NIRSpec and NIRCam data}
\label{s.data}

\gal was selected from the JWST Advanced Extragalactic Deep Survey \citep[JADES;][]{Rieke2020a,Bunker2020,Eisenstein2023}. In our analysis, we use both NIRSpec/MSA spectroscopy \citep{Jakobsen2022, Ferruit2022} and NIRCam imaging data publicly available from JADES \citep{Rieke2023}. The data were selected as part of the "medium\_jwst\_gs" tier (PID:1286).
The source \gal was selected from the data via visual inspection of strong high-z balmer break candidates as part of the identification of high-z quiescent galaxies \citep[specifically for][]{Baker2024c}.
From NIRSpec, we use the prism spectrum (spectral resolution $R=30\text{--}300$), which was observed with 3-shutter slitlets and integrated for 2.2~hours. For the NIRSpec data reduction, see \citet{Bunker2023,D'Eugenio2024}. For the photometry, we use Kron photometry from PSF-matched imaging (point spread function) and the latest data reduction \citep{Eisenstein2023b}.

\section{Spectral features}
\label{s.analysis_spectrum}

\begin{figure}
    \centering
    \includegraphics[width=1\linewidth]{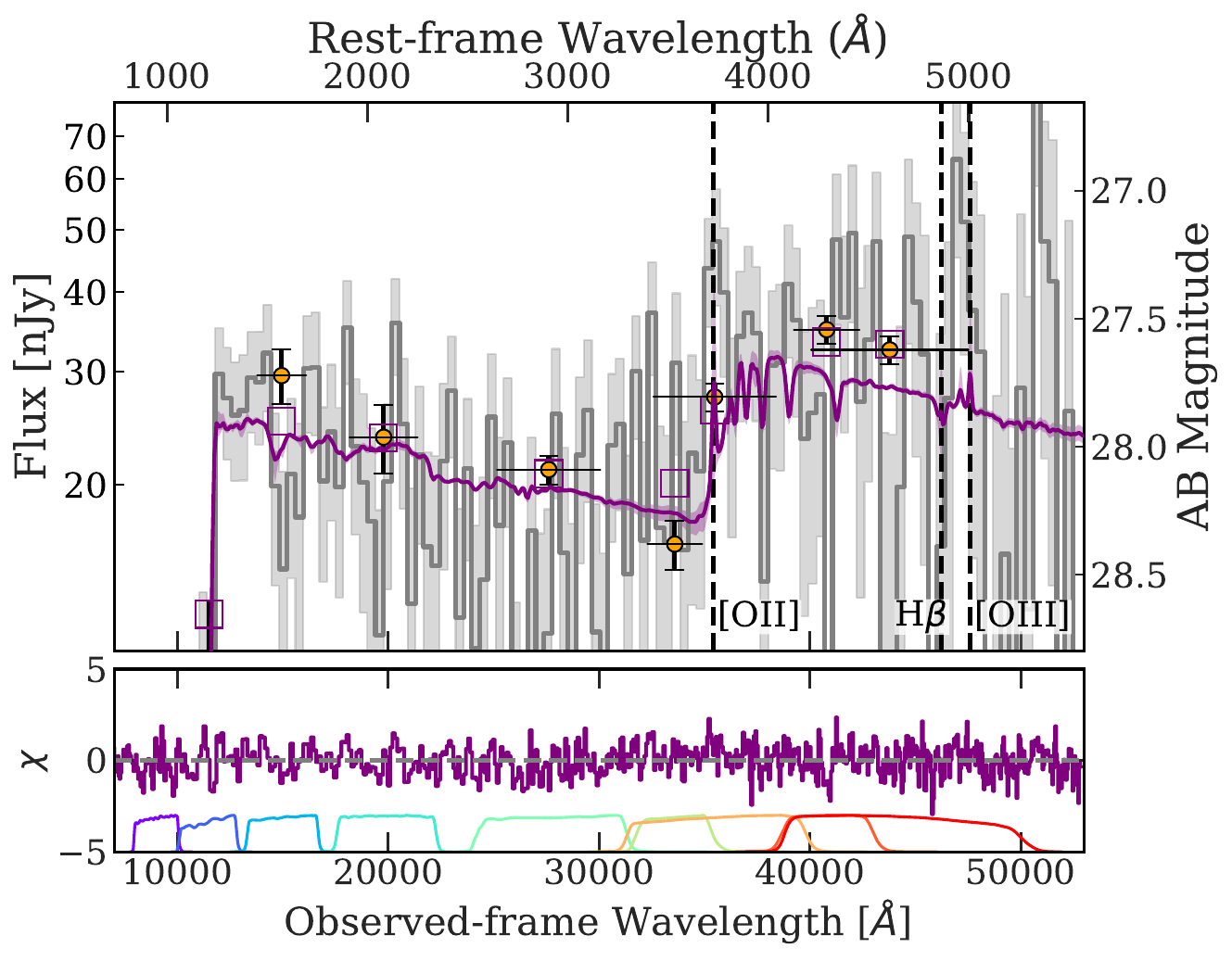}
    \caption{Upper panel: 1D spectrum for \gal in $\rm F_\nu$. This figure highlights the Balmer break in the spectrum. The observed spectrum (grey) has been rebinned to improve visualisation. The best-fit \prospector spectrum and photometry is in purple. The shaded region of the best-fit spectrum shows the 16th and 84th percentiles. The dashed lines correspond to common emission lines. The normalised fit residuals are shown in the bottom panel alongside the NIRCam filter curves. }
    \label{fig:fnu_spec}
\end{figure}

The first aspect that is apparent from the spectrum (Fig.~\ref{fig:Spectrum}) is a lack of strong emission lines and a steep \be slope. A Balmer break is also present, but is more noticeable in $\rm F_\nu$ (see Fig. \ref{fig:fnu_spec}). 
This immediately invites comparison to the mini-quenched galaxies such as in \cite{Looser2024}. These have steep UV slopes, visible Balmer breaks, and no signs of strong emission lines.
In this section, we will explore the spectral features, starting with potential emission lines followed by the \be slope. 

\subsection{Emission-lines}
\label{ss.em_lines}
Although, \gal does appear similar to previously observed "mini-quenched" galaxies, in our case
we also appear to have a tentative detection of \otwolong (hereafter [OII]; 2.9~\sig; P-value<0.002 \footnote{Note the one-tailed P-value without redshift-sweep corrections is appropriate here, because the redshift is fixed by the Lyman$\alpha$ and Balmer breaks, cf.~\citet{Hainline2024}.}). 
This is a complicated emission line to measure the significance of because it is almost coincident with the Balmer break at the prism resolution. 
To accurately model the shape of the continuum underlying \otwo, we use \textsc{ppxf} \citep{Cappellari2017, Cappellari2023}, following the setup of \citet{D'Eugenio2024}. We measure all emission line fluxes from the continuum subtracted spectrum.
 Unlike \otwo, \othreelong (hereafter [OIII]), which is almost always more luminous than \otwo at these redshifts \citep[][]{Cameron2023,Sanders2023,Witten2024}, is not detected (1.2~\sig).

If the \otwo emission was confirmed, this would yield a 3\sig upper limit on O32 of $\log(\rm O32)<0.06$, suggesting a very low ionisation parameter.
We also check for \cthree emission, finding it is undetected with a SNR of 1.2. 

The Balmer lines are undetected, with a 3\sig upper limit for the H$\beta$ flux of $\rm <27.9\times 10^{-20}\;ergs/s/cm^2$. We can then convert this into a 3\sig upper limit on the SFR \citep[assuming the Balmer decrement ratio between H$\alpha$ and H$\beta$ of 2.86 to obtain a H$\alpha$ flux, e.g.][]{Osterbrock1989}. 

It is important to account for sub-solar metallicity within high-z galaxies for this conversion. To do this, we follow the approach of \cite{Shapley2023} who build upon the analysis of \cite{Reddy2018}. We therefore use the \cite{Reddy2018} conversion factor of 
\begin{equation}
    \rm SFR =10^{-41.67}L_{H\alpha}.
\end{equation}
which is appropriate for higher-z lower-metallicity systems.
This gives us a star-formation rate of $\rm <1.3\;M_\odot\; yr^{-1}$.
We can do the same with \othree assuming a [OIII]/H$\alpha $ conversion value of 1.2 from the literature \citep{D'Eugenio2024} which gives us a 3\sig upper limit of $\rm <0.5\;M_\odot\; yr^{-1}$. We also explore using a conversion value of 3.4/4.8 using the emission line ratios of the stacks from \citet{Roberts-Borsani2024} which gives an upper limit of $\rm <0.4\;M_\odot\; yr^{-1}$ from the \othree.
These limits suggests that this galaxy is not highly star-forming. We cannot fully rule out a multi-component scenario with a UV-bright clump plus a heavily obscured star-forming component \citep[see][]{Faisst2024}, because we have no access to sufficiently red wavelengths.
However, in general, the mass range relevant to our galaxy shows no evidence of highly obscured star formation, with all galaxies consistent with little to no dust \citep[e.g.,][]{Sandles2024,McClymont2024}.
We include the main line flux upper limits in Table \ref{tab:properties}.

\subsection{Balmer break}
Fig. \ref{fig:fnu_spec} shows the (rebinned) observed NIRSpec spectrum and best-fit model  spectrum in $F_\nu$. We see that both the observed and model spectrum show evidence for a Balmer break, usually a sign of an older stellar population and strongly linked to the stellar mass and star-formation rate of a galaxy \citep{Wilkins2024}. We measure the strength of the Balmer break following the procedure of \citet{Binggeli2019, Wilkins2024} which consists of the ratios between a blue window from 3400\AA-3600\AA and a red window from 4150\AA to 4250\AA. 
For the observed spectrum we obtain a value of  1.14 $\pm$ 1.18 due to the large errors on the spectrum at these wavelengths and the small window used. Due to this, we use the best-fit prospector value which gives a value of 1.75 $\pm$ 0.02. 
Either one of these breaks is stronger than comparable galaxies at these redshifts \citep{Roberts-Borsani2024, Witten2024} suggesting that \gal shows unusual properties for such a high-z galaxy.
We explore possible stellar population models that can produce this break in Sec. \ref{s.SED_modelling_tensions}.

\subsection{\be slope}
\label{ss.beta_slope}
\begin{figure}
    \centering
    \includegraphics[width=\linewidth]{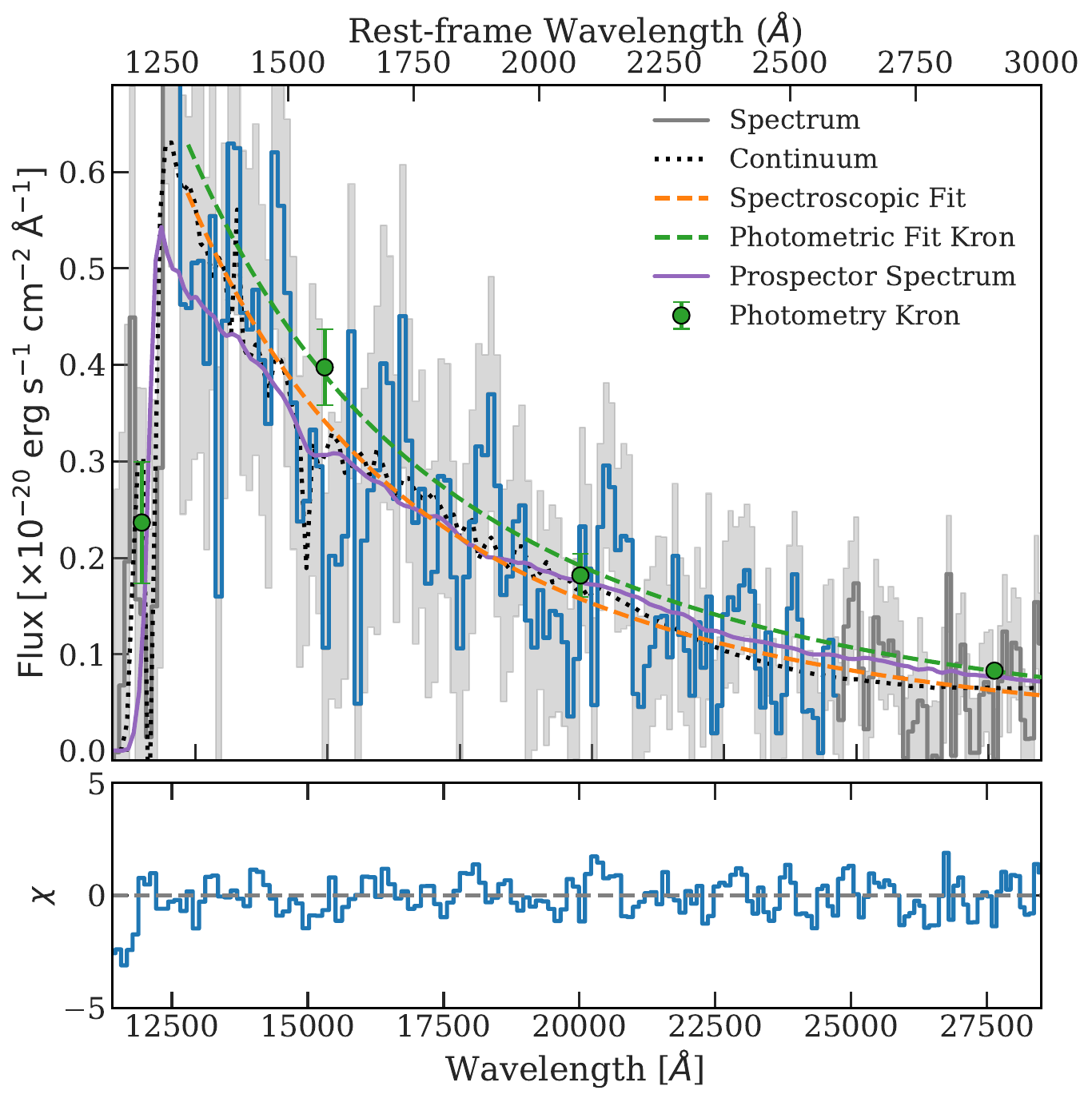}
    \caption{Upper panel: A zoom-in of the \be slope region of the spectrum (grey). The blue region is used for calculating the fiducial beta slope and the orange dashed line is the best-fit fiducial \be slope from the observed spectrum. The Kron Convolved photometry is overplotted as the green points and the best fit \be slope from it is the green dashed lines. 
    The black dotted line is the \textsc{ppxf} best-fit to the continuum emission. We see that the orange spectroscopic \be slope fit accurately traces the \textsc{ppxf} continuum fit despite these being fit completely independently. 
    We also overplot the best-fit prospector spectrum in purple to show that it cannot match the \be slope measured from the photometry or spectroscopy.
    The lower panel shows the $\chi$ values for the spectroscopic \be slope fit compared to the spectrum. We extend these beyond the fitted region to show the goodness of fit even after extrapolation (at least until the Lyman $\alpha$ break). 
    }
    \label{fig:beta_slope}
\end{figure}
Simply from visual inspection of the spectrum, we can clearly see the steep value of the \be slope. 

We fit the $\beta$ slope using the code \textsc{LMFIT} \citep{Newville2014} following multiple different procedures.
First, it is important to note that we detect no significant emission or absorption lines in the spectrum (aside from the tentative \otwo noted earlier on; see Sec \ref{ss.em_lines}). The absence of strong emission lines minimises their possible effect on the determination of the \be slope regardless of the window used. 
We model the \be slope as a standard power-law form
\begin{equation}
    \rm F \propto \lambda^\beta_{UV}
\end{equation}
where we fit for the normalisation and \be slope.
This means we fit the equation
\begin{equation}
    \rm F =\alpha \lambda^{\beta_{UV}}
\end{equation}
to find $\alpha$ (the normalisation) and \be.
Our fiducial approach is to use the window from \cite{Heintz2024}.
This approach corresponds to fitting the flux and errors within a window between rest-frame 1250\AA and 2600\AA. Fig. \ref{fig:beta_slope} shows the window denoted by the blue region. 
This gives us a measured \be slope value of \be=-2.8$\pm$0.2 which is shown in Fig \ref{fig:beta_slope} as the orange dashed line.
We also test using a similar procedure to that of \citet{Roberts-Borsani2024} using a window of 1600\AA to 2800\AA. This is a slightly redder region designed to minimise potential contributions from emission lines. This gives us a value of \be=-$3.0\pm0.3$, i.e. resulting in a bluer slope but still remaining consistent with the other window.
In addition to these two windows, we also explore fitting the blue side of the UV slope to obtain a quantity defined as $\beta^{1550}$ by \cite{Chisholm2022}. This uses a window of 1300\AA to 1800\AA and has been shown to be a predictor of the escape fraction (\fesc). Using this bluer window gives us a value of $\beta^{1550}=-2.6\pm 0.5$, which again remains consistent with our fiducial value (albeit with larger measurement uncertainties due to the reduced window size). Propagating this value and errors forward we can use the relation from \cite{Chisholm2022} to estimate the escape fraction which is given by 
\begin{equation}
    \rm f_{esc}= (1.3\pm0.6)\times10^{-4}\;\times 10^{(-1.22\pm0.1) \beta^{1550}}.
\end{equation}
This gives us \fesc=$0.2^{+0.5}_{-0.1}$, which, while relatively unconstrained, does seem to indicate an escape fraction of ionising photons >10\%.

As an independent check on the \be slope measured from the spectrum, we also fit the photometric bands F150W, F200W, and F277W with a powerlaw. These bands trace approximately the same spectral region as the window in the spectrum. We obtain a \be slope of $-2.5\pm 0.1$. Clearly, this does not fully agree with the spectroscopic measurements, although they remain within around 1\sig. We produce mock photometry from the spectrum and find a small difference between the \be slopes produced with it and those of the spectroscopy, because it is not probing exactly the same regions of the spectrum.

A key aspect to keep in mind is that the NIRSpec microshutter does not probe the entirety of the galaxy. This is made clearer by Fig. \ref{fig:rgb}, which is an RGB image of the galaxy with the open area of the NIRSpec microshutters overplotted. We see that the galaxy appears extended (more details on this in Sec \ref{s.analysis.morphology}) and that it appears to be non-central within the NIRSpec shutter. 

\begin{figure}
    \centering
    \includegraphics[width=\linewidth]{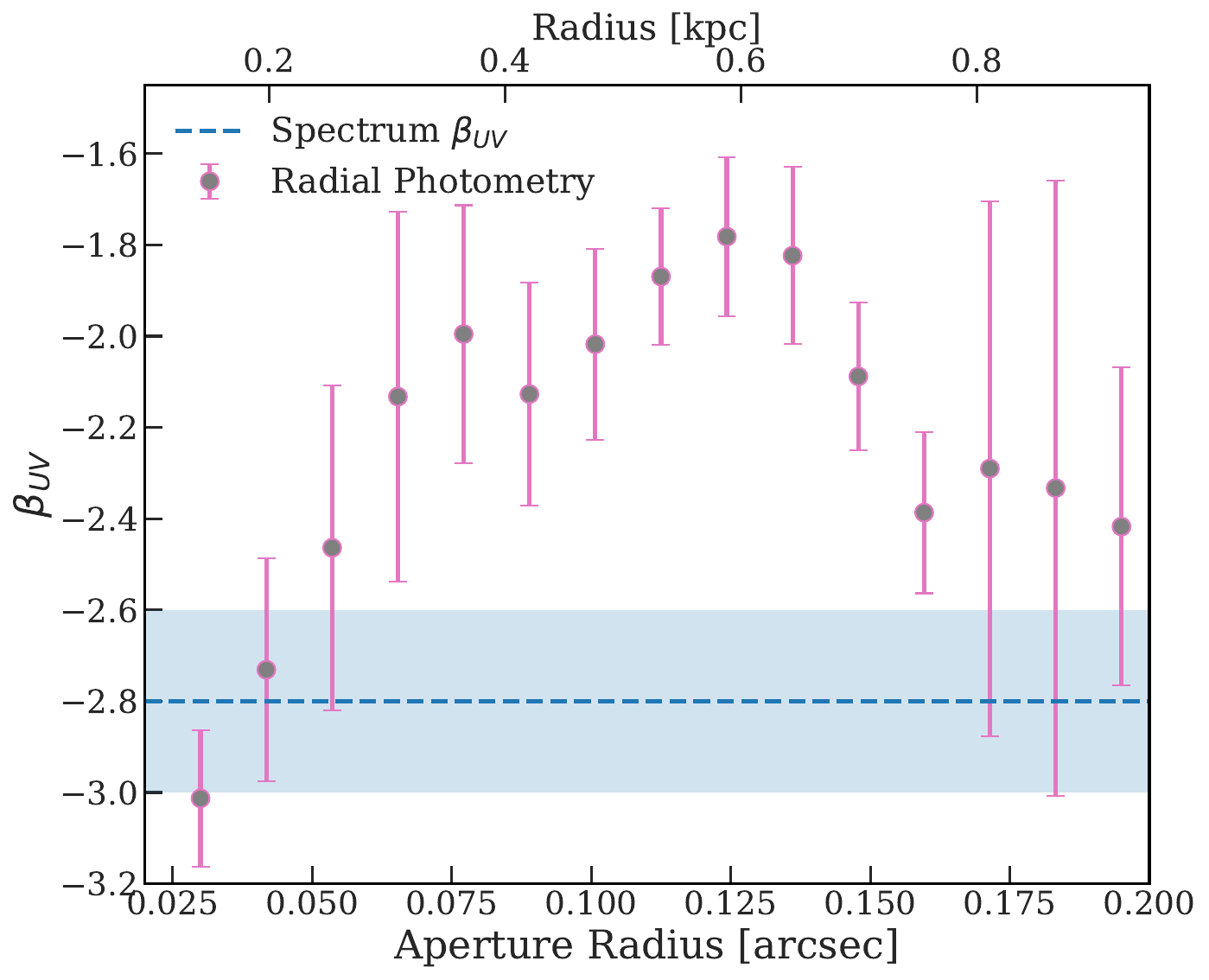}
    \caption{The photometric UV slope (\be) versus aperture radius for circular annuli on the PSF matched imaging. The spectroscopically measured \be slope is overplotted as the dashed blue line for comparison with its error corresponding to the shaded region.}
    \label{fig:radial_beta}
\end{figure}

Using PSF matched photometry, we can explore any spatial dependence of the \be slope. We draw circular annuli around \gal (using the centre obtained via \sersic fitting in Sec \ref{s.analysis.morphology}) and compute the \be slope based on the same three NIRCam bands. The result is shown in Fig. \ref{fig:radial_beta}. We see that depending on the region of the galaxy probed, the \be slope varies from a value of around -3 for the most central region to a value of -2.0 for the outskirts. This helps explain any possible mismatch between the value measured from the spectrum and the photometry. It appears from the photometry alone that it is dependent on the region of the galaxy probed. In the central region, the galaxy appears to have a steep \be slope, possibly corresponding to a region of higher \fesc (given the simultaneous lack of strong emission lines), whilst further away from the centre the \be slope becomes slightly shallower.

We will refer to the \be slope of this galaxy as \be=-2.8$\pm$0.2 throughout this paper, but want to make it clear we are reporting the spectroscopic value, likely corresponding to the particular region of the galaxy falling within the slit, and not the overall galaxy as a whole which appears more consistent with a value of \be=-2.5$\pm$0.1.

This \be value of -2.8 is steep even for high-z galaxies \citep[see e.g.][]{Roberts-Borsani2024} and means that we are seeing little dust reddening combined with multiple other processes, such as little to no nebular continuum emission. For comparison, the mini-quenched galaxy from \cite{Looser2024} has \be = -2.09.

\section{Morphology}
\label{s.analysis.morphology}

\begin{figure}
    \centering
    \includegraphics[width=0.9\columnwidth]{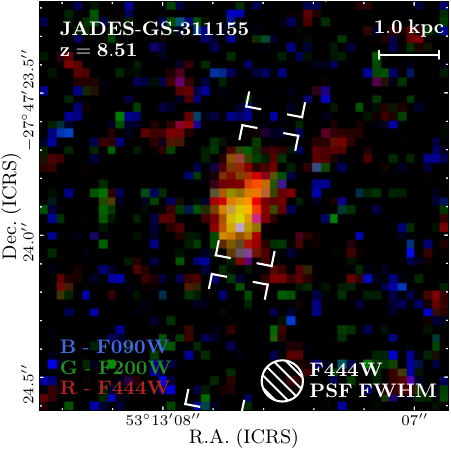}
    \caption{An RGB image of \gal in the F444W-F200W-F150W filters. The NIRSpec slit is overplotted alongside the F444W point-spread-function (PSF) and a 1kpc scale bar. 
    }
    \label{fig:rgb}
\end{figure}

\begin{figure}
    \centering
    \includegraphics[width=\linewidth]{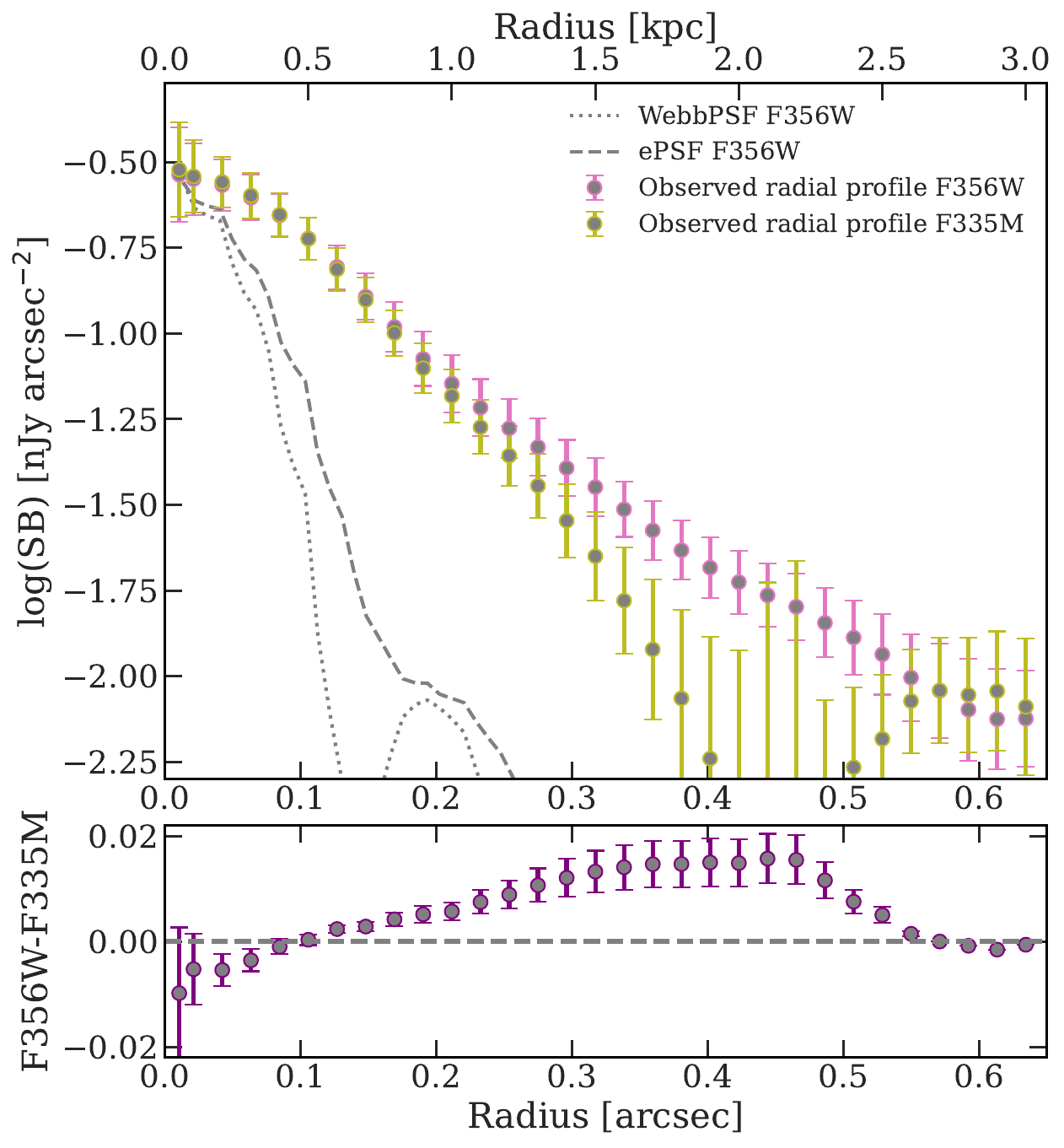}
    \caption{Upper panel: Surface brightness profile of \gal. We can see a deviation between the radial profiles of F356W (containing \otwo) and F335M (which doesn't) at 0.2 to 0.5 arcsec, which is consistent with the expelled weakly ionised gas scenario. Lower panel: the difference between the two radial profiles versus wavelength.}
    \label{fig:surfbright}
\end{figure}

The NIRCam images show a compact source (Fig.~\ref{fig:rgb}) that is clearly resolved, but without obvious structure. We do not find any prominent neighbours within 1.5 arcsec, ruling out this as a clear satellite galaxy. To quantify its morphology, we use \forcepho \citep[Johnson, B. +, in prep;][]{Robertson2023, Baker2024b}, following the setup of \citet{Baker2024b}.

We obtain a \sersic index of 0.98$\pm$0.14, consistent with a disc-like morphology and a half-light radius of 0.05\arcsec$\pm$0.01 -- meaning that the galaxy, whilst small, is still spatially resolved by NIRCam. This corresponds to an effective radius of around 240~pc at z=8.5.

We can explore radial profiles of \gal. The key aspect we want to explore is not only the light distribution, but also the differences in the light distribution between F335M and F356W. With F335M versus F356W we can capture morphological differences between the gas (traced by [OII] in F356W) and the stars (traced by the stellar continuum in F335M).

We therefore measure the surface density profiles for \gal following the procedure in \citet{Baker2024b}. An important aspect to note is that (like in Sec. \ref{ss.beta_slope}) we use the PSF convolved NIRCam imaging where all filters have been convolved to the PSF of the F444W filter. Although this means that we lose information, it does mean that we avoid measuring the differences in the PSF between filters.
We also note that the Balmer break falls between these two filters as well, so some of the difference between the filters can also be ascribed to this. However, we would normally expect the Balmer break to be centred on the galaxy, i.e. the stellar populations, than the possible extended \otwo emission. We cannot rule out the possibility that this extended difference in the radial profiles is down to a Balmer break,  which could be caused by stellar populations far outside of the galaxy. Such a scenario would arguably be even more interesting than extended, low-ionization gas, and would require deep spectroscopy to be confirmed.

Fig. \ref{fig:surfbright} upper shows the radial surface density profile for \gal in the F335M, and F356W filters. The F356W empirical PSF and \textsc{WebbPSF} are overplotted for reference. We see that both radial profiles are much more extended than the PSFs. We can also see that at larger radii the difference in the profiles of F356W and F335M is more pronounced. This is clarified in the bottom panel of Fig. \ref{fig:surfbright}, which indicates the difference in the radial profile between F356W and F335M. We can see within the central region (dominated by the PSF) there is little difference, whereas at larger radii we see a boost in flux in F356W compared to F335M until we reach the noise. 
This is consistent with a scenario whereby the \otwo emission corresponds to gas expelled from the galaxy, but we cannot exclude the possibility that it corresponds to a Balmer break seen within stellar populations stripped from the galaxy.

\section{Prospector and stellar population properties}
\label{s.analysis.sedfitting}

The next stage is to infer the stellar population properties of the galaxy and to uncover its star-formation history (SFH). To do this we utilise full spectro-photometric fitting with the SED fitting code \prospector  \citep{Johnson2021}. For more details on the setup see \citet{Tacchella2022,Baker2024c}.

Our approach models the spectrum and photometry simultaneously in a Bayesian framework. SEDs are produced using the Flexible Stellar Population Synthesis \textsc{FSPS} code \citep{Conroy2009, Conroy2010}. We use the MILES library of spectra \citep{Sanchez-Blazquez2006, Falcon-Barroso2011} with MIST isochrones \citep{Choi2016} and a \citet{Chabrier2003} IMF. Nebular continuum and line emission is included as in \citet{Byler2017}. We perform runs with and without nebular emission to check our results. 
We use a two-component dust model consisting of a foreground dust screen as in \citet{Kriek2013} and extra power-law dust attenuation towards birth clouds \citep[i.e. affecting stars younger than 10~Myr][]{Charlot2000}. However, we note that this makes little difference due to our lack of dust. 
We use two forms of flexible SFH, the "continuity" version from \citet{Leja2019} and the adapted "bursty" prior from \citet{Tacchella2022}. A 2nd order polynomial is used to upscale the spectrum to the photometry.

Our initial modelling attempts could not reproduce the observed UV slope. We deem it unlikely to be a contribution from an AGN continuum, due to the lack of strong emission lines. \footnote{Note that while AGN scenarios displaying Balmer breaks have been proposed \citep{Inayoshi2024}, these require large column densities of gas, which would presumably be accompanied by lower-density gas seen in emission.} In addition, if it were to be an AGN dominating the UV continuum, this would indicate that the UV is dominated by the accretion disc, therefore it would appear as a point source, whereas we find that the galaxy is spatially resolved.
Hence, we explain the steep UV slope as a result of light from young stars escaping their birth clouds. 

Therefore, we also incorporate the free parameter "frac\_obrun" which represents the fraction of `runaway' O- or B-type stars, i.e. not embedded into their birth clouds,  which we use as an \fesc parameter.

Fig. \ref{fig:Spectrum} and Fig. \ref{fig:fnu_spec} purple (upper panels) show the best-fit spectrum and photometry.
We see that the spectrum well reproduces the clear Lyman-$\alpha$ break and the Balmer break (see Fig. \ref{fig:fnu_spec} and Section \ref{s.SED_modelling_tensions}).  
However, the best-fit spectrum has a \be slope of \be=-2.30, which is shallower than that expected from the spectroscopy or photometry (see Fig. \ref{fig:beta_slope}). 
We explore this in more detail in Sec. \ref{s.SED_modelling_tensions}.

\begin{figure*}
    \centering
    \includegraphics[width=1\linewidth]{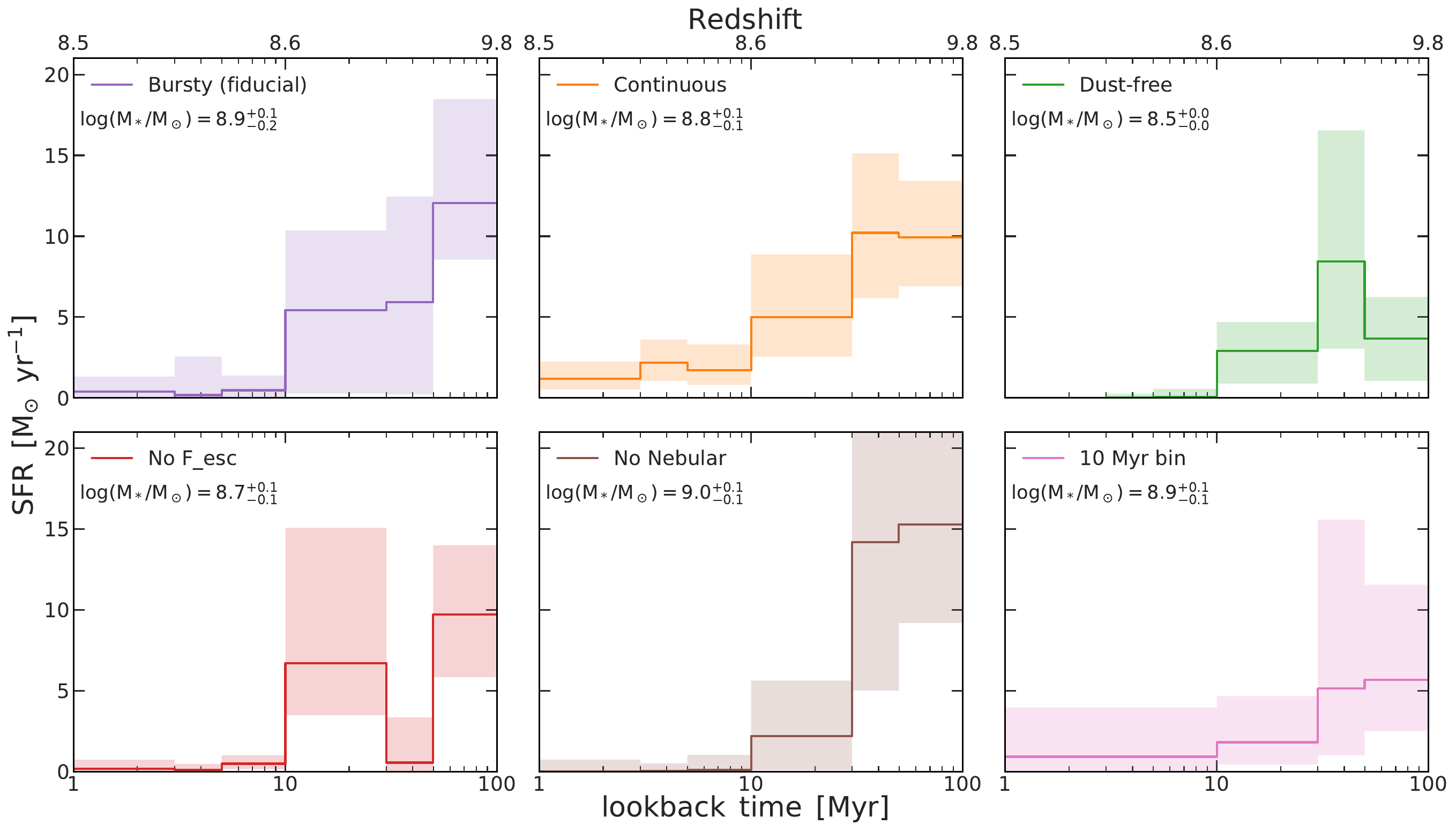}
    \caption{Star-formation rate versus lookback time (redshifts). Comparison of the SFHs from different \prospector runs with varying priors and models. In purple is the fiducial bursty continuity prior run. In orange is the \citet{Leja2019} style continuity prior run. In green is a run with no dust parameters (i.e. dust free) and in red is a run with no \fesc parameter (i.e. \fesc=0). In brown is a run with no nebular component and in pink is a run with the most recent star-formation bin spanning 10Myrs instead of the fiducial set up. Despite the varying models we obtain a consistent interpretation of a rapidly quenched SFH.}
    \label{fig:SFH_comparison}
\end{figure*}

Fig. \ref{fig:SFH_comparison} shows the resulting SFH for a multitude of priors and models. 
Our fiducial model is the upper left model in purple.
We measure a stellar mass of $\rm M_\star=10^{8.9\;\pm 0.1}M_\odot$ for the bursty prior. This still places \gal in the mass range where burstiness is suspected to play a role \citep{Ceverino2018,Looser2024, Lovell2023}.
We find \gal appears to have been forming stars up until the last 10Myr when the SFR finally declined. This rapid truncation is consistent with some form of feedback expelling the gas from the galaxy \citep[e.g.][]{Gelli2023}. 
In addition, the SFH tells us that this galaxy "quenched" within the last 30~Myr ($t_{90}=27^{+26}_{-7}$ Myr) with a formation time of $t_{50}=66^{+7}_{-15}$ Myr (where $t_{90}$ and $t_{50}$ are the lookback times, inferred from the SFH, at which the galaxy formed 90\% and 50\% of its stellar mass respectively).

The other panels in Fig. \ref{fig:SFH_comparison} show the effect of changing the priors of different \prospector runs on our recovered SFHs and stellar masses. Fig. \ref{fig:SFH_comparison} shows the results for varying the SFH prior, dust, \fesc, binning and nebular component.
All runs follow the fiducial run except for one key change.
In orange is exactly the same run but with a continuity SFH prior, i.e. the Student's $t$ distribution scale parameter is set to 0.3. This weights the SFH against large SFR changes between adjacent time bins, thereby biasing the model towards a more continuous SFH. We can see the effects of this prior in the SFH returned, which is a slightly more continuous SFH than for the bursty prior, however, both appear to tell a consistent story.  
The stellar mass returned is $10^{8.8\pm 0.1}\rm M_\odot$, so consistent with the bursty case. 

In addition to the SFH, we can test varying the dust model. We remove any modelling of dust (i.e. saying the galaxy is dust-free). In this case (shown in green) we see a similar result to the bursty run, but with the star-formation (and hence stellar mass) reduced. This shows that without any dust we cannot "hide" any further star-formation within the spectrum, reducing the overall mass and recent SFR of the galaxy. However, this remains consistent with our fiducial model in terms of interpretation, i.e. a recent episode of star-formation in a  burst followed by rapid quenching. 

The next thing to test is the effect of excluding \fesc. We run a version of the fiducial run but with the "frac\_obrun" parameter set to 0 (the default setting in \prospector) -- this SFH is shown in red. We see that this does not appear to change much, only it gives a double peaked SFH, once again highlighting the possible impact of burstiness. 

We then remove entirely the nebular component, meaning that we do not model any emission lines or nebular continuum. This naively, one would expect to make little difference as we see no strong emission lines in the spectrum and the steep UV slope suggests virtually no nebular continuum. We see a very similar SFH to our fiducial run.

The final aspect we try is to use a 10-Myr bin as the last bin for SF prior to observation, as typically assumed in non-parametric SED modelling. The result is shown in pink. We can clearly see that what this does is essentially average over the quenching period of the fiducial run that takes place within those 10Myrs.

It is worth cautioning that in the short timescale regime in the most recent bins, the exact SFH is likely to be highly sensitive to the prescriptions used for modelling extremely young stellar-evolutionary phases, such as Wolf-Rayet stars.

The key results from the SED fitting are shown in Table \ref{tab:properties}.

\section{Escape fraction}
\label{s.escape_fraction}

From the spectroscopy and photometry, we infer a steep $\be=2.8\pm0.2$. Given the known correlation between \be and escape fraction, this steep \be suggests a non-negligible value of \fesc. This is confirmed by the PSF-matched photometry, as arising from the more central regions of the galaxy. Although the neutral inter-galactic medium makes it impossible to measure $\rm f_{esc}$ directly at $z=8.5$, we can use the best-fit value of the "frac\_obrun" parameter in our fiducial \prospector run as a rough estimate of $\rm f_{esc}$. We find a value of $\rm f_{esc}=0.5^{+0.3}_{-0.2}$, although, as obtained from SED fitting, this cannot be particularly constraining, it does seem to support a scenario in which \gal has a high escape fraction of ionising photons, as would be expected from the steep UV slope.  

However, it is important to note that the \fesc obtained from the "frac\_ob" parameter in \prospector assumes no extra dust attenuation in the birth clouds, while dust can
linger around young stars even after most of the gas has been ionized \citep{Charlot2001}. Our dust-attenuation from SED fitting is relatively unconstrained, although there is unlikely to be significant amounts due to the steep value of the UV slope (see Sec. \ref{ss.beta_slope}). In the presence of dust, there are the effects of the Lyman continuum photons interacting with the dust to consider \citep{Tacchella2023b}.  These could be absorbed which could significantly reduce the escape fraction.
Therefore, we caution that this is an estimate and should be considered with the other estimate from the \be slope. 

As mentioned in \ref{ss.beta_slope}, if we calculate a value of \fesc from the \be slope measured using the criteria of \citet{Chisholm2022}, we get a value of \fesc$=0.2^{+0.5}_{-0.1}$ which is consistent with our SED modelling value. Combined, the conclusion from these two methods appears to be that \gal has a non-zero escape fraction, likely >10\% although we cannot constrain this in detail.

We can explore how this compares to other LyC leakers at lower and comparable redshift. Part of what makes \gal so interesting is the mismatch between the \be slope and the emission lines from the spectrum. Steeper \be slopes than -2.8, have been seen previously from spectroscopy \citep{Heintz2024, Saxena2024, Yanagisawa2024}, but on average they are less steep than \gal \citep{Roberts-Borsani2024}. However, those sources observed with steeper \be slopes have booming \othree emission lines and high ionisation parameters \citep[such as the highlighted examples in][]{Saxena2024, Dottorini2024}. In contrast, \gal has no sign of \othree emission, and even more extraordinarily there is the tentative \otwo emission at 2.9\sig which, if confirmed, would imply a very low ionisation parameter.

\begin{figure}
    \centering
    \includegraphics[width=0.95\linewidth]{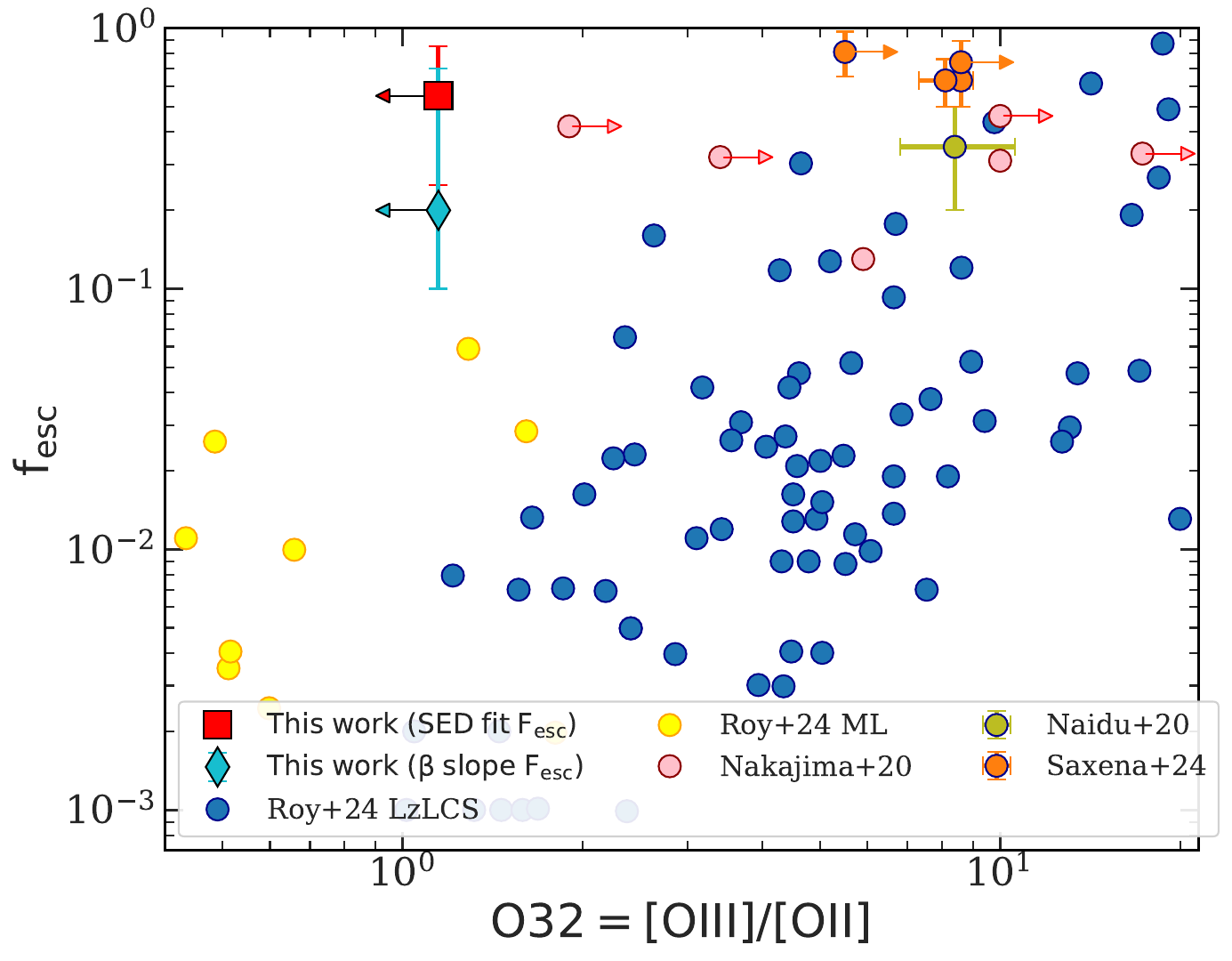}
    \caption{Escape fraction $\rm f_{esc}$ against O32 ratio for a number of different LyC leakers. These include the Low-redshift Lyman Continuum Survey \citep[LzLCS, z=0.2--0.4, blue points][]{Saldana-Lopez2022, Flury2022} and other low-redshift massive starbursts from \citet{Roy2024}. We also plot detections and lower limits (pink points) from \citet{Nakajima2020} and the high \fesc average from \citet{Naidu2022}, all of which are z=2--3. As orange points we plot values from \citet{Saxena2024} (in some cases O32 lower limits) with \fesc calculated from the \be slope via the \citet{Chisholm2022} method. 
    We report our two values as the red square for the SED fit \fesc and the turquoise diamond for the \citet{Chisholm2022} \fesc.
    We see that \gal is clearly in a totally different region of the diagram to all these other galaxies due to the unusual combination of high \fesc and extremely low O32.}
    \label{fig:o32vfesc}
\end{figure}

Fig. \ref{fig:o32vfesc} shows \gal in a diagram of \fesc versus O32 \citep[a strong tracer of ionisation parameter with a secondary dependence on metallicity,][]{Maiolino2019}. The two different escape fractions we measure are denoted by the red square and the light blue diamond, respectively.

Also included are the Low-redshift Lyman Continuum Survey \citep[LzLCS, z=0.2--0.4, blue points][]{Saldana-Lopez2022, Flury2022} and other low-redshift massive starbursts from \citet{Roy2024}. We also plot the detections and lower limits (pink points) from \citet{Nakajima2020} and the high \fesc average from \citet{Naidu2022}, all of which are at z=2--3. As orange points we plot values from \citet{Saxena2024} (z=5--8, and in some cases O32 lower limits) with \fesc calculated from the \be slope via the \citet{Chisholm2022} method. 

As previously mentioned, it is hard to constrain \fesc from the SED fitting due to the "frac\_obrun" parameter in \prospector not taking into account the presence of dust. 
In short "frac\_obrun" in \prospector informs on the ionising photons from OB stars that are outside their birth clouds, but then does not track them through the remainder of the ISM. Therefore, with a dusty ISM, it cannot tell us how many of these actually escape the galaxy.
However, we see clearly that if \gal has an \fesc>10\% (consistent with both SED and \be slope estimates) it is still in a completely different region of this diagram compared to the other observed galaxies.
It is clear that this is due to the highly unusual combination of high \fesc and extremely low O32.

This combination makes \gal (or at least the subregion of \gal probed by the NIRSpec slit) a strong Remnant leaker candidate.

\section{Possible model tensions and caveats}
\label{s.SED_modelling_tensions}

In this section we further explore the results of the SED modelling with regards to possible caveats and tensions.

As shown in Fig.~\ref{fig:fnu_spec}, we see clear evidence for the Balmer break in the spectrum and photometry. 
However, as shown in Fig. \ref{fig:beta_slope} we also see clear evidence for a very steep beta slope from both the spectrum and the photometry.

Our best fit-fiducial \prospector model (and every other model we try) gives us a best-fit \be slope of -2.3. This remains shallower than either the spectrum or the photometry of \gal. This is apparent in all of Figs.~\ref{fig:Spectrum}, \ref{fig:fnu_spec} and~\ref{fig:beta_slope}.
Here we explore why this is so.

\begin{figure*}
    \centering
    \includegraphics[width=0.9\linewidth]{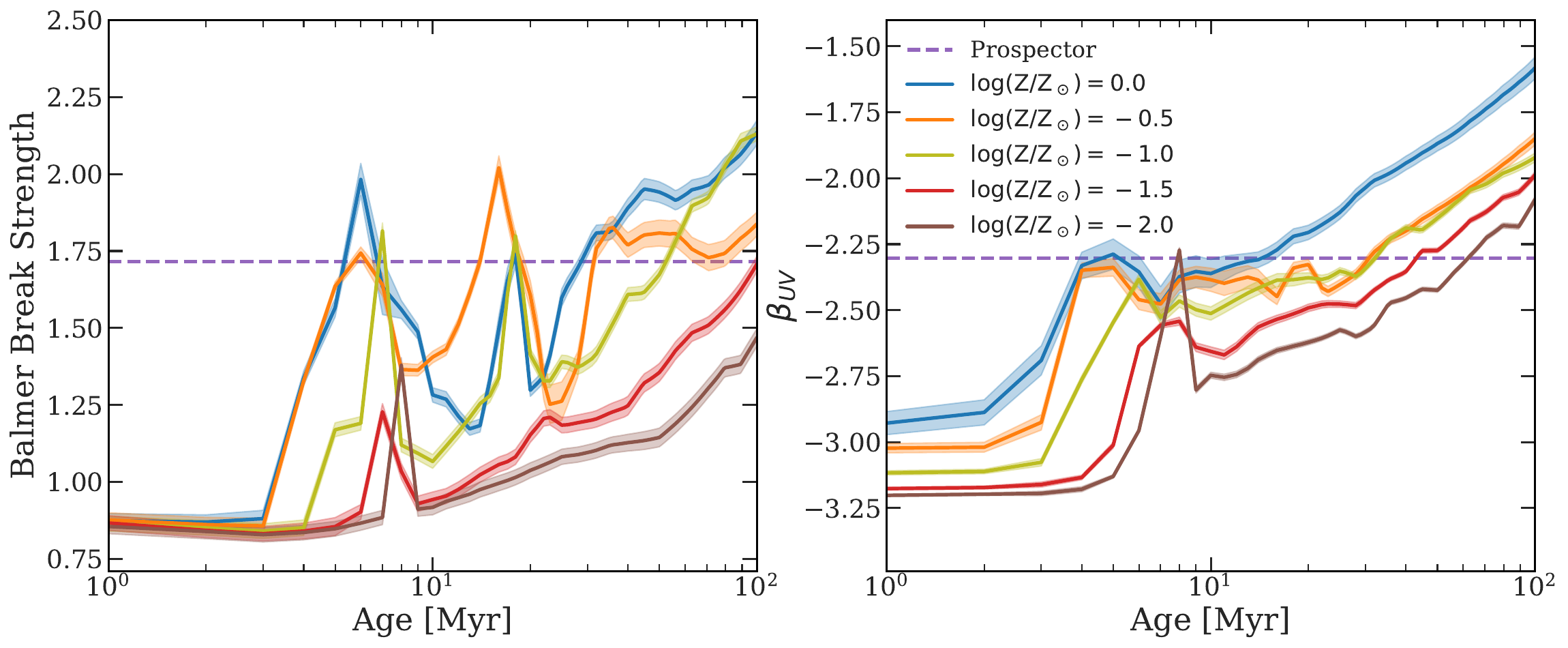}
    \caption{Left: Balmer break strength versus age of stellar population for various tracks of metallicity from mock spectra. Right: Beta slope (\be) versus age of stellar population for various tracks of metallicity from mock spectra. The value measured from our fiducial prospector fit is the horizontal purple dashed line. }
    \label{fig:fsps:balmervage}
\end{figure*}

To do this, we model what type of stellar population is required for this kind of \be slope and Balmer break using the Flexible Stellar Population Synthesis code  \citep[\textsc{FSPS}][]{Conroy2009, Conroy2010}. This enables us to build simple stellar population models for which we can then measure the \be slopes and Balmer breaks and then compare to our observed and best-fit values. The key parameters we want to explore are the stellar ages and metallicities. 
For our stellar population models, in order to match our prospector fitting we again use a \citet{Chabrier2003} IMF, a redshift of z=8.51, and include IGM absorption ( this should not affect our result as this is bluewards of our beta slope measurement). 
We explore an array of ages dating from 1Myr to 100Myr and an array of metallicities from $\rm log(Z/Z_\odot)=-2$ to $\rm log(Z/Z_\odot)=-1$ in steps of 0.5. We expect these quantities to be linked due to the so-called "age, dust metallicity degeneracy" \citep[e.g.][]{Gallazzi2005}.
We caution that in exploring these models we are not including the effect of dust attenuation or nebular continuum, as doing so would over-complexify our models.
These choices are supported by the results of the SED modelling,
which find little dust and no ongoing star formation.
Indeed, any non-zero dust attenuation or nebular continuum emission would flatten \be, making the model mismatch even worse.

Fig. \ref{fig:fsps:balmervage} shows the trends between stellar population age and metallicity and the spectral properties \be and Balmer-break strength. The left-hand panel shows that we can obtain a Balmer break of the strength measured by our fiducial fit (dashed line) with a 5-Myr old stellar population, but we require a high stellar metallicity of $\log Z/Z_\odot>-0.5$. Otherwise we can obtain a Balmer break that strength with a slightly older stellar population with a metallicity of $\log Z/Z_\odot=-1.0$, a lower metallicity than this would require a population aged around 100Myr old. So far, models can reproduce the observations at almost all ages or metallicities, thanks to the onset of colder-atmosphere 
blue giants.
However, once we add constraints from the \be slope (Fig. \ref{fig:fsps:balmervage}, right panel), the model becomes
unable to reproduce the data.
The \be slope tends to prefer a younger or lower metallicity solution compared to the Balmer break, with \be$\leq$-2.8 being consigned almost exclusively to stellar populations younger than 10Myr old. However, such young stellar populations never display a Balmer break as strong as 1.75 without extremely high stellar metallicities.
We can fit either the strong Balmer break with an older and/or more metal-enhanced stellar population or we can fit the strong \be slope with a younger and/or metal poorer stellar population.

All our \prospector models prefer to fit the strong Balmer break over the steep \be slope, thereby favouring an older and/or metal rich stellar population. 

Although \be seems better constrained than the observed Balmer break, in reality the signal-to-noise of the Balmer break feature is not limited to
the narrow regions that are conventionally used to measure the break strength.

Even so, our best fit \prospector model still returns a "frac\_obrun" best fit value of \fesc~"frac\_obrun"=0.5$^{+0.3}_{-0.2}$ suggesting that in the case of \prospector fully reproducing the steepness of the \be slope, the true value for \fesc could be higher.

All together this suggests a couple of interesting possibilities. The first of these is that there may be issues with the underlying models used. This would appear from first glance to be an unusual galaxy \citep[no other clear candidates with similar features have been reported so far][]{D'Eugenio2024, Saxena2024,Kuruvanthodi2024} and may represent an extreme outlier galaxy with peculiar properties (or very short lived period of a galaxies lifetime). 

Secondly, this may call for more flexibility in modelling approaches. At the moment we are limited to modelling a single value for the stellar metallicity of the stellar populations with \prospector and being able to fit two composite stellar populations with different metallicities should be able to resolve this tension. 

Finally, related to modelling two composite stellar populations, this may hint at an underlying multi-component structure for \gal. We could have an older quenched component \citep[similar to the kind of "mini-quenched" galaxies, e.g.][]{Looser2024} along side a component that has shutdown star-formation much more recently within the previous $\sim$10Myr leading to a steep \be slope remaining and a high \fesc.

This strongly motivates further research, both on the modelling side, but also observationally to perform follow-up surveys to find more of these types of galaxies and unveil their properties.

\section{Discussion}
\label{s.discussion}

We have explored \gal, a high-z galaxy at $z=8.5$ with a mass of $M_\star=10^{8.9}\;M_\odot$  with evidence of a steep \be slope of -2.8, a Balmer break, and no evidence of strong emission lines typically associated with star-forming galaxies at $z=8$ \citetext{\citealp{Cameron2023}; H$\beta$ and [OIII], Section~\ref{s.analysis_spectrum}}.

The absence of these lines requires either a lack of very recent star formation, or a high $f_\mathrm{esc}$ (or both).
Morphologically, the galaxy is well described by an exponential light profile with radius 240~pc, indicative of an assembly history dominated by gas accretion and \textit{in-situ} star formation.

The standard SED modelling procedures applied to the spectrum and photometry find that the most likely SFH had a previous burst, followed by a rapid decline. 
There remains the possibility for residual star formation after the decline within the continuity SFH solution, but this is disfavoured by evidence from the spectrum such as the low upper limits on the SFR obtained from the undetected emission lines. Qualitatively, this solution is still consistent with our fiducial interpretation. 
These models, with no escape of ionising photons, do not reproduce the steep observed \be, suggesting a modelling tension between the steep UV continuum, lack of emission lines, and evidence of a Balmer break. This was explored further in Section \ref{s.SED_modelling_tensions} finding that the best-fit models struggle to reproduce simultaneously both the steep \be slope and the strong Balmer break.
By adding a free escape fraction, the best-fit model favours "frac\_ob"$\sim$\fesc=$0.5^{+0.3}_{-0.2}$, which can efficiently suppress nebular emission while maintaining a steep \be. The recent SFR could be higher in these cases, yet the presence of a Balmer break and high \fesc also limit the number of extremely young stars present in the galaxy. This is because if they were present, the unattenuated light from these young stars would erase the Balmer break \citep[e.g.][]{Witten2024}. This model, therefore, still predicts a decline in the very recent SFR, on timescales of 10~Myr.

Fig. \ref{fig:o32vfesc} shows \gal clearly occupies a totally different part of the diagram compared to these other strong LyC leakers. This suggests that it presents a compelling example of unusual (at least more locally) processes at play in the high-z Universe. We do, however, caution that the \otwo emission line used to calculate the O32 upper limit is only tentatively detected and deeper spectroscopy is required to confirm or refute this feature.

The high \fesc of \gal combined with the steep \be slope and likely low-ionisation parameter suggest it is a strong Remnant Leaker candidate (which is also supported by the clear $\rm SFR_{10Myr}/SFR_{100Myr}<1$). These are interesting as, despite having high \fesc, they are unlikely to be significant contributors to reionisation \citep{Katz2023}. From \prospector, we find a value of $\rm \log(\xi_{ion}/[Hz/erg])=24.5^{+0.8}_{-0.9}$ (accounting for \fesc) with $\rm \log(n_{ion}/[s^{-1}])=53.1^{+0.5}_{-0.8}$ suggesting this is likely to be true for \gal as well, with both of these values significantly below average galaxies at these redshifts \citep{Simmonds2024}. However, this depends on the number density of these Remnant Leaker sources and how their properties change when they are in a burst phase. With enough sources and during a burst phase with increased efficiency, they may still be considerable contributors to reionisation. Another possibility is that they are a short-lived dormant phase of the kind of bursty leakers seen more commonly. 
Deep follow-up spectroscopy of larger numbers of Remnant Leaker galaxies is needed to establish this and to explore their number densities and typical properties.

Fascinatingly, the formation time of \gal is around $\sim$70Myr and the quenching time is $\sim$ 30Myr suggesting that \gal is both recently formed and rapidly quenched. This suggests that we are seeing a recently formed galaxy in its first quenching episode after a burst of star-formation. Although it appears to have been rapidly quenched, the spectrum of \gal appears totally different from that of more massive high-z quiescent galaxies, such as the z=7.3 galaxy from \citet{Weibel2024}, which suggests they may have different quenching processes. Indeed, \gal appears to be much more consistent with the kind of mini-quenched galaxy observed in \citet{Looser2024} than the traditionally quiescent high-z galaxies \citep{Carnall2024, Baker2024c, deGraaff2024}. 
Despite the lack of star formation when observed, it is still UV-luminous with $\rm M_{UV}\sim-19.5$~mag.

Morphologically, the galaxy is fairly compact, with indications of a more extended structure in F356W than in F335M. Given that F356W contains both the Balmer break and [OII] and F335M covers neither feature, this morphological difference could indicate either a stronger Balmer break in the outskirts or a nebular emission that is more extended than the stellar emission. 

A fascinating hypothesis is that the extended nebular emission may trace a diffuse, low-ionisation gas nebula.

Although being similar to the \citet{Looser2024} "mini-quenched" galaxy, based on some lower-z metrics, \gal would not necessarily classify as quenched or (depending on definitions) even "mini-quenched" (although definitions vary for this phenomenon). This is due to its steep \be slope and the aforementioned tensions in modelling this alongside the Balmer break (see Sec. \ref{s.SED_modelling_tensions} for more details).
In spite of this, \gal is clearly not actively forming stars in any kind of burst phase when observed; rather, this has happened in its past. In light of this, we refer to this galaxy as "rapidly quenched", and suggest it may be a more distant version of the "mini-quenched" galaxies. It may very well be a progenitor to these kind of "mini-quenched" galaxies, where we observe it closer to its quenching time than seen previously \citep[][]{Looser2024}, hence the persistence of the steep blue \be slope.

A natural question is what caused the sharp decline in star formation for \gal? Based on the SFHs the quenching process would have had to have occurred within the last ~30 Myr after a short burst of star formation. This rules out slow-quenching mechanisms typically seen at lower redshifts, such as starvation \citep{Peng2015, Trussler2020, Baker2024}. 
Without any obvious neighbour or evidence of recent interaction, it seems that environment effects or mergers cannot explain the decline in SFR.

Given the short inferred timescales, it is most likely some form of ejective feedback, such as AGN outflows or radiation-driven winds \citep{Gelli2023}. SNe are also a possibility as their timescale is $\sim$ 30Myr \citep{Gelli2023}, although they may lack sufficient energy to quench \gal \citep[][]{Gelli2024a} . These mechanisms would all be consistent with the tentative evidence for ejected gas from the galaxy causing the observed \otwo emission, and the wide-band filter excess at larger radii.

\gal is in the mass range where we can expect star-formation to possibly reignite after fresh accretion of gas from the cosmic web in some form of rejuvenation \citep{Witten2024}. This could of course perhaps be reignited via re-accretion of the recently expelled gas.

Intriguingly, we find a tentative detection of \otwo (2.9\sig), which would imply that any gas still present in \gal must have very low ionisation ($\log \rm O32 < 0.06$). 
How then do we explain this? \otwo is a low ionisation line, and we clearly see no evidence of strong (or detectable) \othree emission with a low upper limit on the ionisation parameter. This suggests that there may be weakly ionised gas in or around the galaxy. One scenario for explaining this is that the galaxy has expelled gas during the start of quenching and that this gas is still being ionised at a greater distance, hence the low ionisation parameter and, therefore, the emission of \otwo without high-ionisation lines such as \othree. Another option is that \otwo is excited by (slow) shocks in the outflow. An additional option is that this gas is simply cooling or has been affected by shocks.

A further possibility for many of these tensions is a multi-component structure for \gal. We already see some tentative elements of this in the radial \be slope measurements of Sec. \ref{ss.beta_slope}. 
In this scenario, it would be possible to have an older quenched component \citep[very similar to ][]{Looser2024} in one dominant component, whilst also having another high \fesc more recently star-forming component. This could also explain the tensions seen in Section \ref{s.SED_modelling_tensions} which would arise due to the necessity of modelling both components together with a single value for the stellar metallicity.
This possibility would remain consistent with our findings.

\gal appears consistent with the attenuation-free model (AFM) of \citep{Ferrara2023, Ferrara2024}. With its lack of dust attenuation, extremely blue \be slope, mass range and rapidly quenched nature it fulfils all the necessary criteria. In addition, the SFH of \gal even satisfies the sSFR criteria ($\rm sSFR\geq$20 Gyr$^{-1}$) required for the AFM at the burst time. This makes \gal a possible "blue monster", the overly luminous high-z galaxies observed \citep{Ferrara2024}. The radiation driven outflow required to expel the dust would describe many properties of \gal. The outflow would expel both dust and gas leading to rapid quenching, blue colours and a high escape fraction suppressing nebular line emission. The tentative extended \otwo emission would then correspond to the gas expelled by the outflow.
To confirm this scenario, possibly addressing the issue of over-luminous galaxies, it is crucial to obtain deeper observations, particularly targeting the emission lines. Higher-quality data are essential to confirm our tentative detection of \otwo and to investigate the physical mechanisms driving the nebular emission and possible outflow.

\begin{table}
      \caption[]{Properties of \gal.}
         \label{tab:properties}
         \begin{tabular}{c|c}
            \hline
            \noalign{\smallskip}
                   ${\rm \gal}$ & ID: 60311155 \\
            \noalign{\smallskip}
            \hline
            \noalign{\smallskip}

            RA &  53.218802 \\

            Dec &   -27.789978\\
            
            \begin{tabular}{@{}c@{}} [OII] \\ $[10^{-20}\rm erg/s/cm^2]$ \end{tabular} &  $<27.9$
            \\
            
            \begin{tabular}{@{}c@{}} $\rm [OIII]$ \\ $[10^{-20}\rm erg/s/cm^2]$ \end{tabular} & $<32.09$
            \\

            \begin{tabular}{@{}c@{}} $\rm H\beta$ \\ $[10^{-20}\rm erg/s/cm^2]$ \end{tabular} & $<9.9$
            \\
            
            \be & $-2.8\pm0.2$
            \\

            \begin{tabular}{@{}c@{}} $r_e$ \\ $[\arcsec]$ \end{tabular} & $0.05\pm 0.01$
            \\

            n & $0.98\pm 0.14$
            \\

            \begin{tabular}{@{}c@{}} $\rm SFR_{5}$ \\ $[\rm M_\odot/yr]$ \end{tabular} & $0.3^{+1.4}_{-0.3}$
            \\

            \begin{tabular}{@{}c@{}} $\rm SFR_{10}$ \\ $[\rm M_\odot/yr]$ \end{tabular} & $0.6^{+0.8}_{-0.1}$
            \\

            \begin{tabular}{@{}c@{}} $\rm SFR_{100}$ \\ $[\rm M_\odot/yr]$ \end{tabular} & $8.9^{+3.1}_{-3}$
            \\

            $\rm \log(M_\star/M_\odot)$ & $8.9^{+0.1}_{-0.2}$
            \\

            $f_{\rm esc}$ & >$10\%$
            \\
            \noalign{\smallskip}
            \hline
         \end{tabular}
         \tablefoot{The RA, DEC, half-light radius and \sersic index are obtained via the \forcepho best-fit model.
         The line fluxes correspond to the 3\sig upper limits. Only \otwo is detected at 2.9 \sig. 
         The star-formation rates, stellar mass and escape fraction are obtained from SED modelling.}
   \end{table}

\section{Conclusions}
\label{s.conclusions}

   \begin{enumerate}
    \item We report the discovery of \gal, a rapidly quenched, redshift 8.5 (\Mstar=$\rm 10^{8.9}M_\odot$) galaxy with a steep spectroscopic \be slope of -2.8 and a clear Balmer break.
    \item As measured from photometry, we find that the \be slope varies radially from around -3.0 in the centre to -2.2 in the outskirts, possibly suggesting regional variations in the escape fraction, ionisation, or dust properties.
    \item We find that the galaxy is well-described by an exponential light profile and is spatially resolved by NIRCam with a half-light radius of around 240pc, suggesting a disc-like nature, disfavouring recent major mergers.
    \item NIRCam photometry shows a positive radial gradient in F356W-F335M colour, suggesting spatially diffuse extended \otwo emission.
    \item In the spectrum, we find a tentative detection of \otwo at 2.9\sig and no other sign of emission lines, possibly signifying a very low ionisation parameter ($\log \mathrm{O32}<0.06$).
    \item From both SED fitting and the high \be slope we show that the galaxy appears to be a strong "Remnant Leaker" candidate, with a nominal $\rm f_{esc}$ of >10\%.
    \item We find that \gal appears to have rapidly quenched (<30Myr) and that the red part of the spectrum is consistent with other high-z "mini-quenched" galaxies, albeit at a higher redshift and with a much steeper \be slope.
    \item We show that the data can be explained by a scenario in which extremely rapid feedback has expelled the gas and halted star formation, leaving a diffuse cloud of weakly ionised gas.
    \item \gal also appears consistent with an attenuation-free model caused by radiative outflows of dust and gas, which would neatly describe its properties of a very blue \be slope, high escape fraction, rapid quenching and lack of strong line emission. 
    %\item \gal and any other similar galaxies are crucial to understand because of their importance to both galaxy quenching and reionisation.
   \end{enumerate}
\gal presents clear challenges to our understanding of galaxy evolutionary timescales. The combination of the steep blue \be slope and the clear Balmer break provide strong challenges to existing modelling frameworks (or hint at an underlying multi-component structure). In galaxy terms, \gal shows extreme galaxy features (steep spectroscopic UV slope, strong Balmer break) yet no strong emission lines, making it a convincing Remnant Leaker candidate. It also appears consistent with an attenuation free model with radiation driven outflows. All of this greatly motivates further research into these curious in-between systems and their role in high-z galaxy evolution.

\begin{acknowledgements}
We thank A.~Ferrara for his valuable insights.
WMB gratefully acknowledges support from DARK via the DARK Fellowship. This work was supported by a research grant (VIL54489) from VILLUM FONDEN.
WMB, FDE, RM, CS, and GCJ acknowledge support by the Science and Technology Facilities Council (STFC), ERC Advanced Grant 695671 ``QUENCH'', and by the UKRI Frontier Research grant RISEandFALL.
RM also acknowledges funding from a research professorship from the Royal Society.
AJB and JC acknowledge funding from the “FirstGalaxies” Advanced Grant from the European Research Council (ERC) under the European Union’s Horizon 2020 research and innovation program (Grant agreement No. 789056). 
ST acknowledges support by the Royal Society Research Grant G125142. JW gratefully acknowledges support from the Cosmic Dawn Center through the DAWN Fellowship. The Cosmic Dawn Center (DAWN) is funded by the Danish National Research Foundation under grant No. 140.
SA acknowledges grant PID2021-127718NB-I00 funded by the Spanish Ministry of Science and Innovation/State Agency of Research (MICIN/AEI/ 10.13039/501100011033)
S.C acknowledges support by European Union’s HE ERC Starting Grant No. 101040227 - WINGS.
ECL acknowledges support of an STFC Webb Fellowship (ST/W001438/1). 
BER acknowledges support from the NIRCam Science Team contract to the University of Arizona, NAS5-02015, and JWST Program 3215. The authors acknowledge use of the lux supercomputer at UC Santa Cruz, funded by NSF MRI grant AST 1828315.
The research of CCW is supported by NOIRLab, which is managed by the Association of Universities for Research in Astronomy (AURA) under a cooperative agreement with the National Science Foundation. 
YZ acknowledges support from a JWST/NIRCam contract to the University of Arizona NAS5-02015. We acknowledge use of \textsc{Astropy} \citep{AstropyCollaboration2013}, \prospector \citep{Johnson2021}, \textsc{FSPS} \citep{Conroy2009, Conroy2010}, \textsc{ForcePho} (B. Johnson, in prep.), \textsc{SEDPY} \citep{Johnson2019}, \textsc{numpy} \citep{harris2020array}, \textsc{matplotlib} \citep{Hunter:2007}, \textsc{TOPCAT} \citep{Taylor2005}, and \textsc{LMFIT} \citep{Newville2014}.
This work is based [in part] on observations made with the NASA/ESA/CSA James Webb Space Telescope. The data were obtained from the Mikulski Archive for Space Telescopes at the Space Telescope Science Institute, which is operated by the Association of Universities for Research in Astronomy, Inc., under NASA contract NAS 5-03127 for JWST. These observations are associated with program\#1286. The data is from the JADES survey with the associated MAST link doi:10.17909/8tdj-8n28 and is available at https://archive.stsci.edu/hlsp/jades/.
\end{acknowledgements}


\bibliographystyle{aa}
\bibliography{refs}

\begin{appendix}

\onecolumn
\section{Photometry}

Table \ref{tab:photometry} contains the observed Kron Convolved photometry of the 9 bands used in this work for \gal and the respective errors in nJy. For more details on the photometry see \citet{Eisenstein2023, Rieke2023}.

\begin{table}[h!]
      \caption[]{Photometry of \gal.}
         \label{tab:photometry}
         \begin{tabular}{ccc}
            \hline
            \noalign{\smallskip}
                  Filter & Kron Convolved Photometry (nJy) \\
            \noalign{\smallskip}
            \hline
            \noalign{\smallskip}

F090W & -1.88 ± 3.38\\ 
F115W & 10.41 ± 2.77\\
F150W & 29.59 ± 2.91\\
F200W & 23.71 ± 2.90\\
F277W & 21.10 ± 1.10 \\
F335M & 16.15 ± 1.41  \\
F356W & 27.40 ± 0.93  \\
F410M & 34.91 ± 1.58  \\
F444W & 32.46 ± 1.29  \\

            \noalign{\smallskip}
            \hline
         \end{tabular}
   \end{table}

\begin{figure}[h!]
\centering
    \includegraphics[width=1\columnwidth]{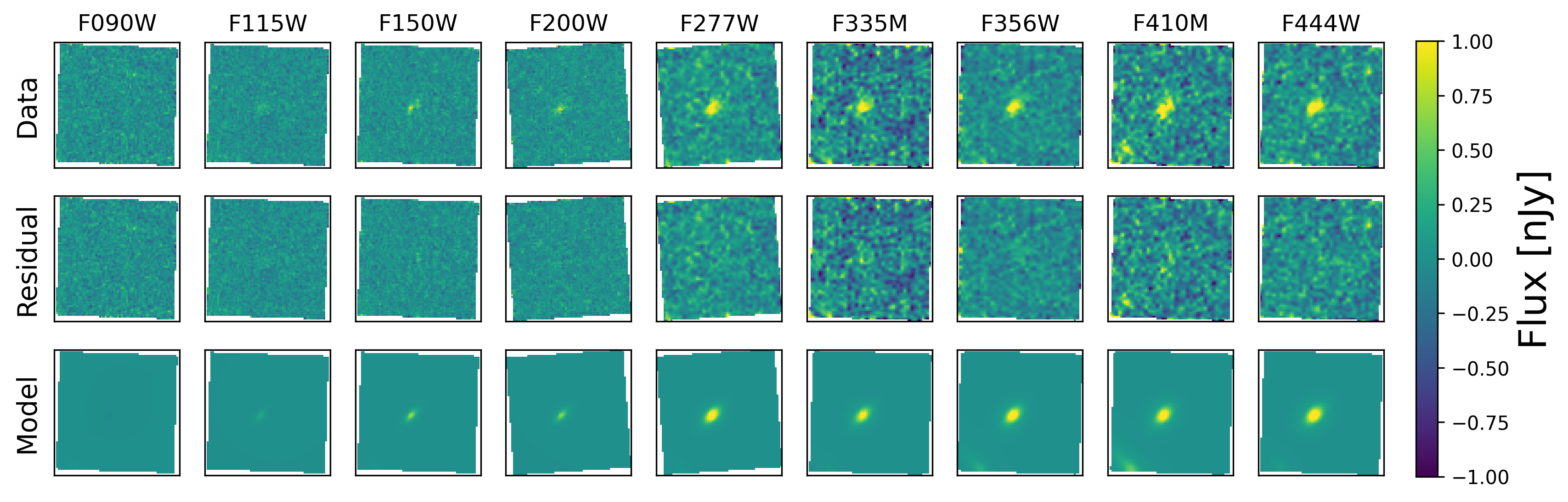}
    \caption{\forcepho fit for \gal in all the NIRCam photometric bands. Upper row shows the data, middle the residual and lower the best-fit model.}
    \label{fig:forcepho}
    
\end{figure}

Figure \ref{fig:forcepho} shows the data, residual and model stacked images for the \forcepho fits of \gal in all the bands fit. We can clearly see that the best-fit model reproduces the data without evidence of significant residuals. For more details on the \forcepho set up see \citet{Baker2024b}.

\end{appendix}
\end{document}